\documentclass[jcp, longbibliography, amsmath,amssymb,  preprint,citeautoscript, superscriptaddress, showkeys]{revtex4-1}
\usepackage{graphicx} 
\usepackage{footmisc} 
\usepackage{bm} 
\usepackage{setspace}
\setcitestyle{super} 
\usepackage{verbatim} 
\usepackage[sort&compress]{natbib}
\usepackage[]{natbib}
\usepackage{multirow}
\usepackage{lineno}
\usepackage{amsmath}
\renewcommand{\thesection}{\arabic{section}} 

\usepackage{rotating}
\usepackage{xfrac} 
\usepackage{hyperref}
\usepackage{cleveref} 
\usepackage{upgreek}
\usepackage{array}
\usepackage{movie15}
\usepackage{appendix}
\usepackage{amsmath}
\newcolumntype{P}[1]{>{\centering\arraybackslash}p{#1}}

\graphicspath{{images_1/}}
\begin{document}
\title{The Desorption Rate at Liquid-Solid Interface}

\author{Krishna Jaiswal}
\affiliation{Department of Chemical Engineering, Indian Institute of Technology, Kanpur, Uttar Pradesh 208016, India} 

\author{Horia Metiu}
\affiliation{Department of Chemistry and Biochemistry, University of California, Santa Barbara, California 93106, USA}
\email{metiu@ucsb.edu}

\author{Vishal Agarwal}
\affiliation{Department of Chemical Engineering and Sustainable Energy Engineering, Indian Institute of Technology, Kanpur, Uttar Pradesh 208016, India} 

\email{vagarwal@iitk.ac.in}


\begin{abstract}
We use a simple generic model to study the desorption of atoms from a solid surface in contact with a liquid, by using a combination of Monte Carlo and molecular dynamics simulations. The behavior of the system depends on two parameters: the strength $\epsilon_{LS}$ of the solid-liquid interaction energy and the strength $\epsilon_{LL}$ of the liquid-liquid interaction energy.  The contact with the solid surface modifies the structure of the adjacent liquid.  Depending on the magnitude of the parameters $\epsilon_{LS}$ and $\epsilon_{LL}$  the density of the liquid oscillates with the distance from the surface for as far as five atomic layers.  For other values of the parameters the density of the liquid near the surface is much lower than that of the bulk liquid, a process sometimes called dewetting.  To describe the desorption rate we have determined the number $N(t)$ of atoms that were adsorbed initially and are still adsorbed at time $t$.  The average of this quantity decays exponentially and this allows us to define a desorption rate coefficient $k_{d}$.  We found that $k_{­d}$ satisfies the Arrhenius relation. The logarithm of the pre-exponential is a linear function of the activation energy (the so-called compensation effect).  We have  also examined two approximate mean-field like theories of the rate constant: one calculates the activation free energy for desorption; the other uses transition state theory applied to the potential of mean force.  Both methods fail to reproduce the exact desorption rate constant $k_{d}$.  We show that this failure is due to the presence of substantial recrossing of the barrier (which introduces errors in transition state theory) and the presence of large fluctuations in the desorption rate of individual molecules whose effect is not captured by mean-field theories.

\end{abstract}
\maketitle
\section{Introduction}
The desorption of molecules from a solid surface immersed in a liquid is an elementary process of interest to catalysis and electrochemistry.  It is frequently assumed that the desorption rate is given by  
\begin{equation}\label{eqn:1}
\frac{dN_{a}(t)}{dt}=-kN_{a}(t),
\end{equation}
where $N_{a}(t)$ is the number of adsorbed molecules per unit area and $k$ is the desorption rate coefficient.  Measuring the temporal evolution of the surface concentration of an adsorbate is possible, in principle, by second harmonic generation or sum-frequency generation spectroscopy, \cite{Lambert2005, Eisenthal1992} but we are not aware of work in which such measurements have been used to monitor how $N_a(t)$ changes with time.  The rate of desorption can also be determined by electrochemical methods \cite{bard2001scanning}  but unfortunately, in the mathematical description of these experiments the desorption rate is a boundary condition (giving the desorption flux at the surface) for the diffusion equation describing the movement of the molecule away from the surface.  This makes the desorption rate difficult to measure because it is entangled with diffusion.  The diffusion coefficient for the material near the surface is different from the diffusion coefficient in the bulk; it depends on the distance from the surface and this dependence is unknown.  Finally, conclusive experimental proof that Eqn.\ \ref{eqn:1} is valid is complicated by the presence  of various defects (kinks, steps, atom vacancies) on the solid surface to which molecules bind with different energies.  Unless the measurements are made under conditions that insure surface cleanliness and uniformity,\cite{wieckowski1999interfacial}  the surface is likely to be defective  and the measured desorption rate has contributions from all defect sites.  

	These complications are avoided in atomic level simulations.   The simulations performed here use a single-crystal solid slab whose surface is free of defects, in contact with a  thick film of liquid.  To study the simplest possible system, the solid and the liquid consists of atoms interacting through van der Waals potentials.  This choice of potential makes it possible to relate the properties of the system to the energy scales in the system, namely the strength $\epsilon_{LL}$ of liquid-liquid interaction and the strength $\epsilon_{LS}$ of the solid-liquid interactions.  

In the first step of each calculation we equilibrate the system by performing a Monte Carlo (MC) simulation to generate equilibrium structures for the system. To study the kinetics of desorption we perform molecular dynamics (MD) calculations with the initial positions generated by Monte Carlo and with the initial velocities derived from a Maxwell distribution.  We denote $N_a(0)$ by  the number of molecules located initially at the surface and $N_a(t)$ by the number of those molecules that are still on the surface at time $t$.  For one simulation (one MC  equilibration and one MD run) $N_a(t)$ is a jagged, descending, staircase-like function.  However, an average over many MC and MD simulations will produce a smooth function which we denote $\langle Na(t)\rangle_{\text{MC + MD}}$.   If a simple phenomenological rate equation of the form used in Eqn.\ \ref{eqn:1} is valid we expect that
\begin{equation}\label{eqn:2}
\langle N_{a}(t) \rangle_{\text{MC+MD}}=\langle N_{a}(0) \rangle_{\text{MC}} \exp\left[-\langle k \rangle t\right],
\end{equation}
where the symbol $\langle k \rangle$ is used to suggest that this is a  mean rate constant.  Our simulations show that Eqn.\ \ref{eqn:2}  is valid regardless of the magnitude of the interactions in the system or of the temperature.  One of the goals of our simulations is to verify whether Eqn.\ \ref{eqn:2} is valid.   Given the fact that it cannot be verified experimentally it is of interest to examine, through simulations, whether Eqn.\ \ref{eqn:2}, which is equivalent to Eqn.\ \ref{eqn:1},  is correct.  The simulations presented here show that it is.  

The binding energy of one liquid atom to the solid (in the absence of the rest of the liquid atoms) is equal to 28-38 kJ/mol.  The desorption of such an atom from the surface into the vacuum takes a very long time and cannot be studied with the methods used here. To calculate the desorption rate one must use a methodology specialized for dealing with rare events.  We find that this is not the case for desorption at the solid-liquid interface:  the liquid-liquid interaction shortens the desorption time considerably.  Without this effect our simulations could not have followed the evolution of the system until $Na(t)$ is very close to zero.

Since we have found that Eqn.\ \ref{eqn:2} is correct, we can determine, through simulations, the desorption rate constant at different temperatures.  One expects (but one does not know for sure) that the dependence of temperature is given by Arrhenius formula and the present simulations confirm this. This raises a number of interesting questions.  1. Is the transition state theory applicable to this system? We expect that the answer is no, because the diffusional motion of the molecules will cause a large number of recrossings of the dividing surface. 2. Desorption at a solid-liquid interface is very different from that at a solid-gas interface.  The desorbing molecule must overcome the attraction energy of the solid and this is helped by the attraction exerted by the molecules of the liquid.  However, to leave the surface, the adsorbed atom must spend energy to make room for itself in the liquid. This means that the desorption energy and the desorption rate depend on both $\epsilon_{LS}$ and $\epsilon_{LL}$ in a rather complicated way.  Attempts to analyze desorption of molecules from an electrode, in an electrochemical cell, by using desorption energies measured in ultra-high vacuum (or in gas phase) are qualitatively incorrect.  However, the binding energy of a molecule on a solid  in vacuum is qualitatively useful if one studies the desorption of the same molecule, into the same liquid, from a variety of solids.  It is then likely that the desorption energies into vacuum are correlated to the desorption energy into the liquid, but we do not know for sure. 3. One can calculate, in simulations, the potential of mean force for removal of an adsorbed molecule from the surface into the bulk of the liquid.  An interesting question to ask is what is the connection of the  barrier in the potential of mean force to the activation energy determined from the Arrhenius plot of the rate constant?  

The simulations allow us to investigate the desorption trajectory of individual atoms. The time it takes a specific atom to desorb depends on its particular environment and it can be quite different from the mean desorption-time.  The simulations allow us to study the distribution of the desorption-time.  The Monte Carlo simulations give information about the structural changes in the liquid in the vicinity of the surface.  We find that when $\epsilon_{LL}$ is small, compared to $\epsilon_{LS}$, the molecules near the surface tend to acquire a solid like, layered structure; because of this, the diffusion constant depends on the distance from the surface and is anisotropic. When $\epsilon_{LL}$ is smaller than $\epsilon_{LS}$ the surface undergoes dewetting: the density of the liquid near the surface is smaller than the bulk density of the liquid. In extreme cases vacuum layer is present near the surface.     
  
\section{Methodology}
\subsection{The Model}

We want to examine the simplest possible model and therefore both solid and the liquid consist of atoms.  The solid  is a slab of a face-centered, close packed solid having a (100) surface.  The slab thickness is 6 atomic layers. We used the unit cell parameter of 3.615 \AA\ corresponding to that of a FCC copper crystal.  All calculations were performed in a supercell containing 6$\times$6 surface unit cells.  The atoms in the solid layer furthest from the liquid were held fixed, those in the other layers were allowed to move and adjust to the liquid environment.  The liquid layer contained 520 atoms and formed a film of roughly thirteen atomic layers. The film was thick enough for the density in the middle to reach the liquid bulk density.  The liquid-vapor interface was unaffected by the presence of the solid.   The thickness of the vacuum layer in contact with the surface of the liquid was $\sim$ 100 \AA.   Periodic boundary conditions were used in all three directions. A picture of the system is shown in Fig.\ \ref{fig:geom}. 

The solid atoms interact with each other through a Morse potential
\begin{equation}
\ V_{M}(\vert \vec{r}_{i,j} \vert)=\ D_{e} \exp[-2 \alpha(\vert  \vec{r}_{i,j} \vert - r_e)] - 2 \exp[-\alpha(\vert  \vec{r}_{i,j} \vert - r_e)],
\end{equation}
where $\vert \vec{r}_{i,j} \vert$ is the distance between atoms located at  $\vec{r}_i$ and $\vec{r}_j$. The parameters $\alpha$ and $r_e$ are given in Table \ref{tab:parm}. The order-of-magnitude of the atomization energy calculated with this potential is comparable to that of metals.  

For the solid-liquid and liquid-liquid interactions we used a Lenard-Jones potential\cite{Jones1924} 
\begin{equation}
\ V(\vert \vec{r}_{i,j} \vert)=\ 4 \epsilon \left[ \left(\frac{\sigma}{\vec{r}_{i,j}} \right)^{12} - \left(\frac{\sigma}{\vec{r}_{i,j}} \right)^{6} \right],
\end{equation}
where $\epsilon$ is the parameter defining interaction strength; in the potential describing the liquid-liquid interaction we denote $\epsilon_{LL}$;  in the case of liquid-solid interaction we use $\epsilon_{LS}$. The interatomic potential is set equal to zero when the distance between atoms is larger than 12  \AA.  We have performed simulations with different values for $\epsilon_{LL}$ and $\epsilon_{LS}$ and  kept the same value for $\sigma$ in all simulations.

\subsection{Simulation Methodology}

For Monte Carlo (MC) simulations, the random numbers were generated using the Mersenne-Twister algorithm.\cite{Saito2008}  Each Monte Carlo move  was performed by selecting an atom at random and displacing it, in a random direction, by a random distance between -0.25 \AA\ and 0.25 \AA.  These displacements were accepted or rejected using the Metropolis algorithm.\cite{Metropolis1953} 

The molecular dynamics simulations were performed by integrating Newton’s equations of motion with the velocity-Verlet algorithm \cite{Swope1982} with a time step of 1.0 femtosecond.  The initial velocities were drawn from the Maxwell-Boltzmann distribution using the Box-Muller method \cite{Box1958,Press1992} and the initial positions were generated by Monte Carlo.   The temperature was kept constant by  using the Andersen thermostat.\cite{Andersen1980} 

The manner in which the density of the liquid varies with the distance from the solid surface was calculated by dividing the space occupied by the liquid into slabs (bins) parallel to the solid surface, having a width of 0.01 \AA.   We follow one atom and monitor the number of times $n_{­i}$ in which the atom was in the bin $i$.  This gives us a histogram of the number of visits of each bin, which is equal to the liquid number-density as a function of the distance to the surface.

The potential of mean force (PMF) $w(\xi$) along the $z$-direction is 
\begin{equation}
\exp[-w(\xi)/k_{B}T] =\alpha\frac{\int \exp[-V(\Gamma)/k_{B}T] \delta(\xi-z)d\Gamma}{\int \exp[-V(\Gamma)/k_{B}T]d\Gamma},
\end{equation}
where $\alpha$ is a constant having units of length; its presence makes the right-hand-side of the equation dimensionless.  The symbol $\Gamma$ is shorthand for $\vec{r}_1$, $\vec{r}_2$, ... , $\vec{r}_N$ where $\vec{r}_i$ is the position of the atom labelled $i$ and $N$ is the total number of atoms.  The probability of finding a liquid atom at a distance $z$ from the surface is
\begin{equation}
P(z)dz=\frac{\exp[-w(z)/k_{B}T]}{\alpha}dz.
\end{equation}
 
We compute the PMF, by using Monte Carlo simulations. We divide the space occupied by the liquid into slabs (bins) parallel to the surface.  The width $\Delta_b$ of each slab is taken small enough so that the PMF is constant inside the slab. The probability that a liquid atom is present in the slab $i$ is
\begin{equation}
P(z)\Delta_b=\frac{n_i}{N_{mc}}
\end{equation}

$n_i$ is calculated by monitoring one atom during a very long Monte Carlo run in which all the atoms are moved.  We call a `MC sweep' a calculation in which all atoms were moved (successfully or not) once and denote by $N_{MC}$ the number of sweeps. We perform a very large number of sweeps and record the number of times $n_i$ that the atom was present is the bin $i$. 
 
	A Metropolis type  implementation of a MC simulation is unable to sample high energy configurations. Therefore, we use the multiple-window umbrella sampling method \cite{Torrie1974} and combine the umbrella windows with the weighted histogram analysis method (WHAM).\cite{Kumar1992}  A few details about the implementation of the method are given in the Appendix.  

The activation free energy $\Delta A^{\dagger}$ is calculated by using 
\begin{equation}\label{eqn:8}
\Delta A^{\dagger}=-k_{B}T \ln\left[\frac{\Delta_b \exp[-w(z_d)/k_{B}T]}{\int \exp[-w(z_d)/k_BT]\chi(z-z_d)dz}\right],
\end{equation}
where $z_d$ is the distance between the dividing surface and the solid surface and  $\chi\left(z-z_{d}\right) =1$ if $z < z_d$ and $zero$ otherwise.   The factor $\Delta_b$ makes the numerator dimensionless and it is present because we have approximated Dirac delta function with a rectangular unit-impulse function of width $\Delta_b$ and height $1/\Delta_b$. 
 
The rate of desorption was calculated by molecular dynamics by using the following procedure.  We start one molecular dynamics calculation by using Monte Carlo simulation to generate an equilibrium configuration for the atoms in the system.   This configuration provides initial positions for the molecular dynamics run.   We identify the atoms initially adsorbed at the solid surface, record  their initial number $N_a(0)$ and `tag' them so that we can follow their subsequent evolution.  We then give each atom in this Monte Carlo configuration, velocities chosen from a Maxwell-Boltzmann distribution.  This generates the initial conditions used to  solve Newton's equation for all atoms in the system.  The time step in this calculation was 1 femtosecond.  The temperature was maintained constant by using Andersen's thermostat.  During the molecular dynamics  we monitor the tagged atoms and record the number $N_a(t)$  of the liquid atoms that are still near the surface at time $t$.  We performed such a calculation 100 times, to calculate the ensemble average $\langle N_a(t)\rangle$ of the number of atoms still on the surface at time $t$.  We found that $\langle N_a(t)\rangle$ fitted very well by an exponential, which allowed us to define a desorption rate constant $\langle k_d$ $\rangle$ through
	\begin{equation}
\langle N_{a}(t) \rangle=\langle N_{a}(0) \rangle \exp\left[-\langle k_d \rangle t\right].
\end{equation}
There was no change in the rate coefficient $\langle k_d$ $\rangle$ on increasing the number of trajectories by 100.

\section{Results}

\subsection{Liquid density as a function of the distance to the surface}
\begin{figure}[!h] 
\includegraphics[scale=1]{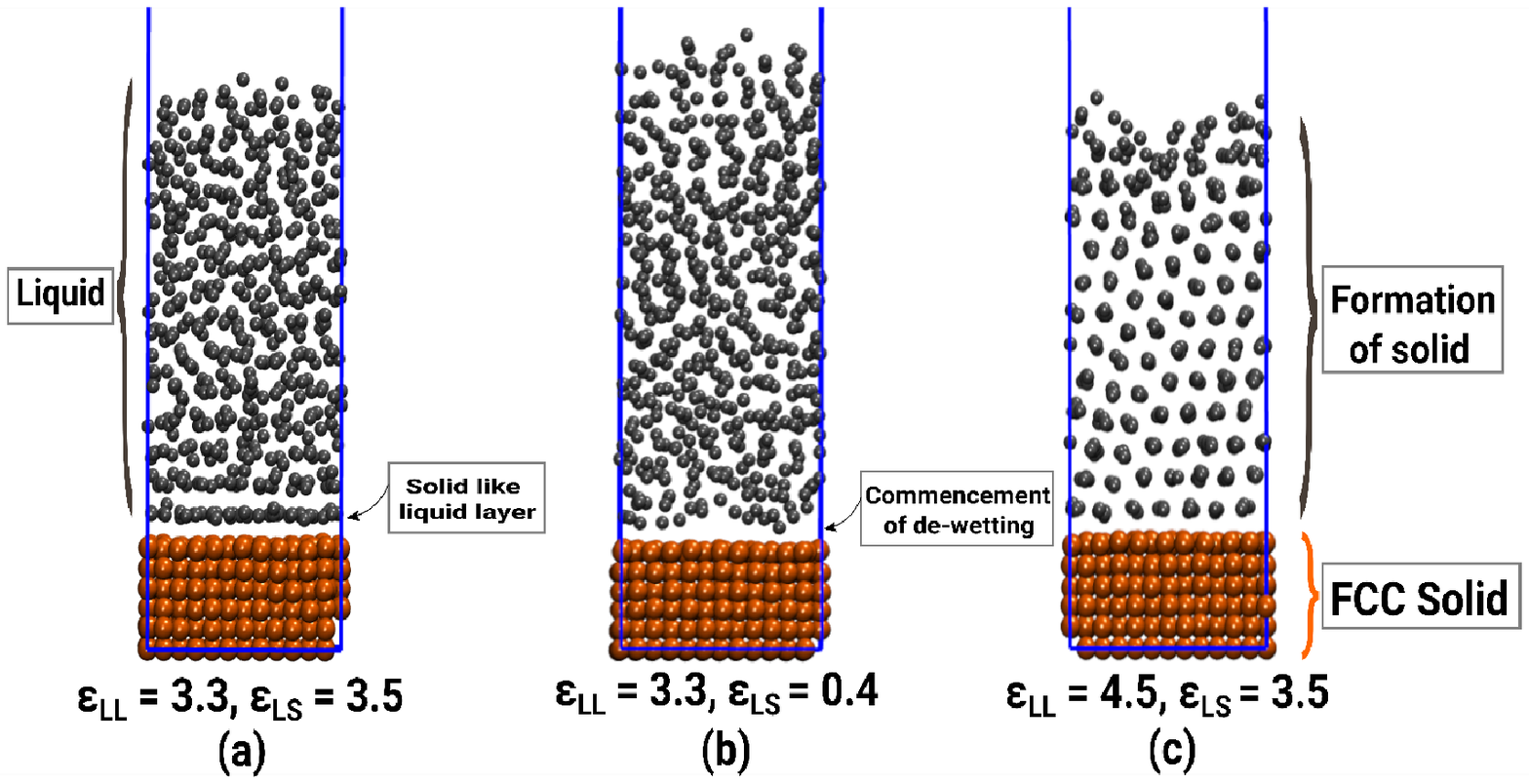}
\caption{Side view of the equilibrated simulation cell at different values of Lennard-Jones parameter $\epsilon$. The subscripts `L' and `S' denotes liquid and solid atoms, respectively. $\epsilon_\text{LS}$ is the Lennard-Jones parameter for the pair-wise interaction between the liquid and solid atoms. The values of $\epsilon$ are in kJ/mol. The solid atoms are shown using brown color and the liquid atoms using gray color. We show boundary of the simulation cell using solid blue lines.}
\label{fig:geom}
\end{figure}

\begin{figure}[!h] 
\includegraphics[scale=0.8]{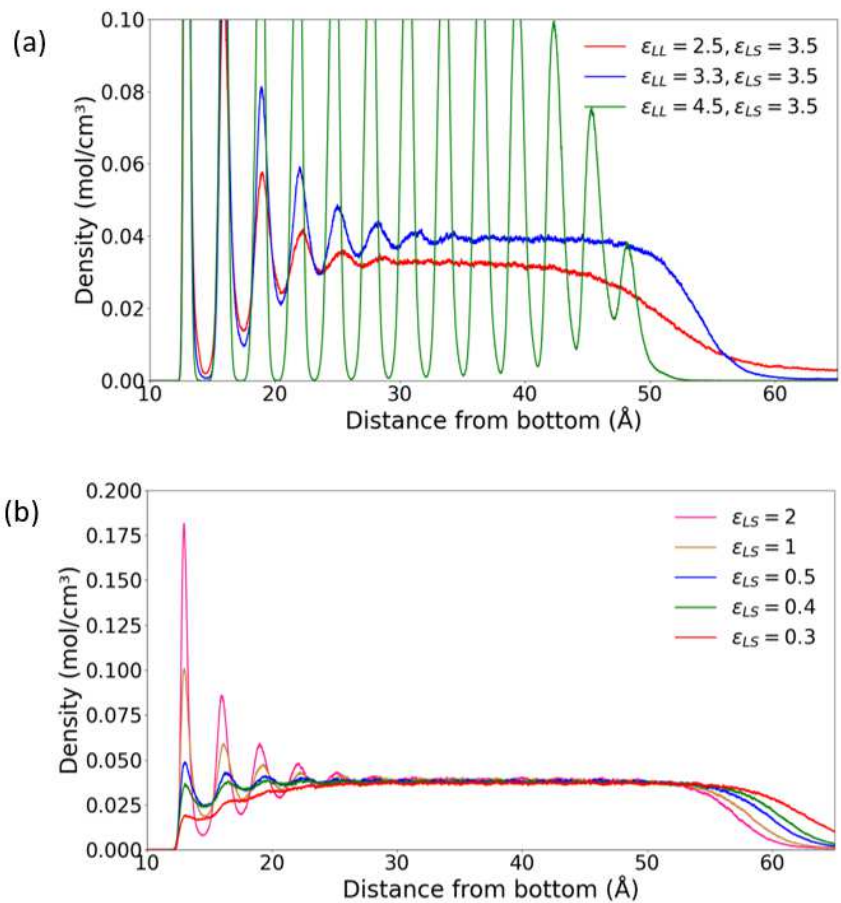}
\caption{Variation of liquid density along the direction perpendicular to the solid surface. The distance is computed from the bottom-most layer of the solid slab, which is kept fixed in our simulations. The solid slab is nearly 11 \AA\ thick, and therefore, the liquid has zero density in that range. Average density of the fluid on the solid surface on \textbf{a)} varying the strength of liquid-liquid interaction at a fixed solid-liquid interaction strength ($\epsilon_{\text{LS}} = 3.5$ kJ/mol) \textbf{b)} varying the strength of solid-liquid interaction at a fixed liquid-liquid interaction strength ($\epsilon_{\text{LL}} = 3.3$ kJ/mol) . The interaction strength is controlled by the Lennard-Jones parameter $\epsilon$, which are shown in kJ/mol. }
\label{fig:densityvar1}
\end{figure}

Fig.\ \ref{fig:densityvar1}a shows how the liquid density varies with the distance to the solid surface, for several values of the liquid-liquid interaction $\epsilon_{LL}$ and $\epsilon_{LS} = 3.5$ kJ/mol at 300 K.   The green curve in Fig.\ \ref{fig:densityvar1}a corresponds to the highest value of $\epsilon_{LL}$, for which the liquid froze and formed a well ordered solid.  The density at the solid-vacuum interface goes to zero gradually which indicates that the solid-vacuum interface is rough, with roughness extending to three monolayers.  A snapshot of the system under these conditions is shown in Fig.\ \ref{fig:geom}c.   The red and blue curves correspond to smaller liquid-liquid interactions.  The flat region at large distance from the surface is the density of the bulk liquid.  The oscillations seen at smaller distances from the surface indicate the formation of   ``solid-like'' layers in the liquid near the interface.  A snapshot of the system  under these condition is shown Fig.\ \ref{fig:geom}a.  The larger the liquid-liquid interaction, the larger the number of solid-like layers induced in the liquid by the presence of the solid surface.  

The change in the liquid-liquid interaction also affects the liquid-vacuum interface: when $\epsilon_{LL}$ is smaller the interface is more diffuse.  When $\epsilon_{LL} = 2.5$ kJ/mol we observe a few molecules in the vacuum; the evaporation of the liquid becomes significant.  
 
Fig.\ \ref{fig:densityvar1}b shows the density dependence on the distance from the surface for a variety of values of $\epsilon_{LS}$  and $\epsilon_{LL} = 3.3$ kJ/mol. The red curve, which corresponds to the smallest value of $\epsilon_{LS}$  ($\epsilon_{LS}  = 0.3$ kJ/mol), the density in the neighborhood of the surface becomes smaller than the density of the bulk liquid, and its dependence on the distance is almost monotonic (very slight oscillations).  This happens because the asymmetry of the forces felt by a molecule near the surface:  larger $\epsilon_{LL}$ than $\epsilon_{LS}$  means that the molecules are more strongly attracted towards the liquid side.  In the limit of larger $\epsilon_{LL}$ a complete dewetting is observed: a ``monolayer'' of vacuum separates the solid from the liquid.  This is similar to the effect of hydrophobicity at a protein-liquid interface.  

More details about the dependence of liquid density of the distance to the surface are given in Figs.\ S1 and S2. 
 
These structural changes in the liquid near the surface have qualitative implications for catalysis.  Since diffusion in solids is much slower than in liquids, the presence of solid-like layers near the solid surface will affect the diffusion of the reactants to the surface and the diffusion of products away from the surface.  

\subsection{The Potential of Mean Force and $\Delta A^\dagger$}

\begin{figure}[!h] 
\includegraphics[scale=0.9]{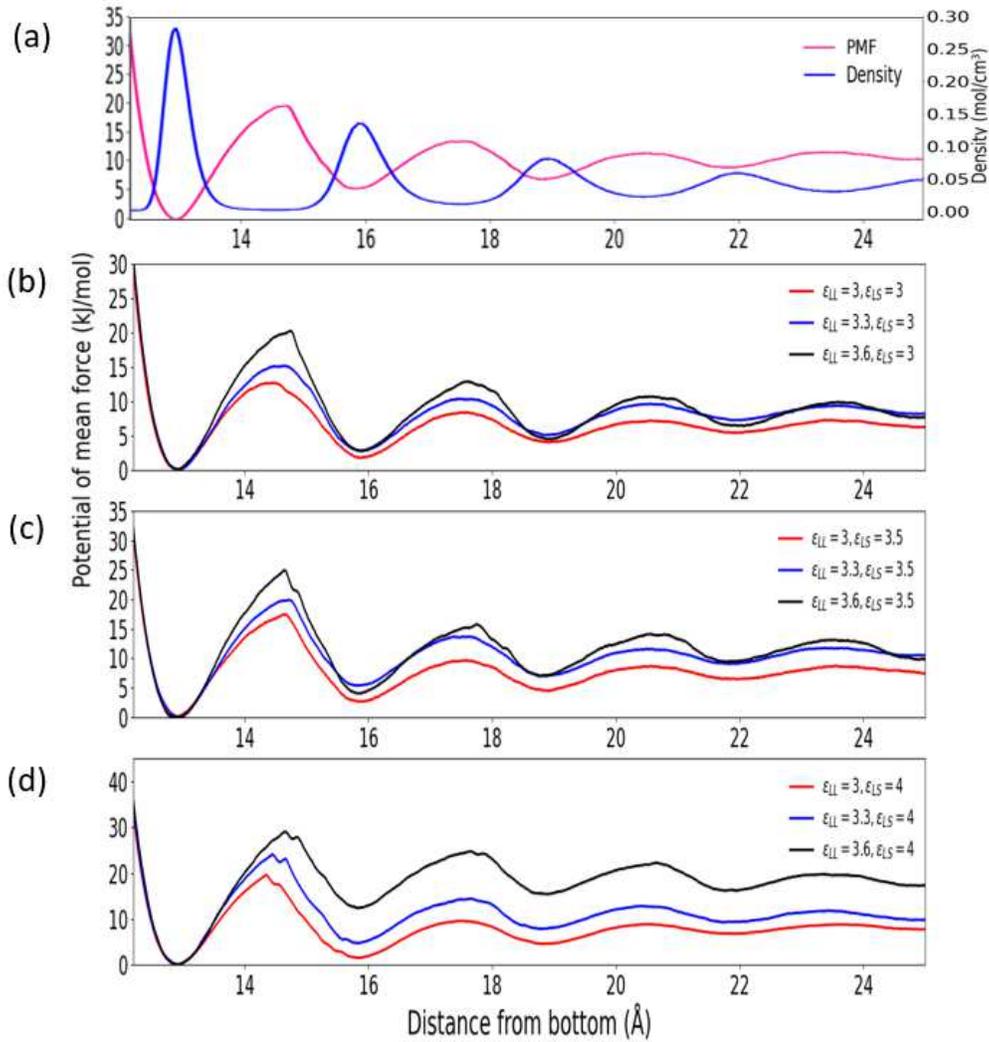}
\caption{\textbf{a)} Potential of mean force and density plotted as function of distance from the solid surface. Lennard-Jones parameters used for these plots are $\epsilon_{\text{LL}} = 3.3$ kJ/mol and   $\epsilon_{\text{LS}} = 3.5$ kJ/mol. \textbf{b)}, \textbf{c)} and \textbf{d)} Potential of mean force as a function of distance from the surface. We show variation of potential of mean force with different liquid-liquid and solid-liquid interaction strength. The values of $\epsilon$ are in kJ/mol.}
\label{fig:pmf2}
\end{figure}

Fig.\ \ref{fig:pmf2} shows  plots of the potential of mean force in the liquid layer, as a function of the distance $z$ perpendicular to the solid surface, for several choices of $\epsilon_{LL}$ and $\epsilon_{LS}$ . The distance $z$ is measured from the rigid solid layer further from the solid-liquid interface;  the liquid film starts at $z = 11$ \AA.  As expected, the potential of mean force has minima at distances for which the density has maxima and becomes constant in the bulk of the liquid film.  The small kinks observed near the maxima are artefacts that will disappear on decreasing the bin width. 
	
	We focus on the free-energy barrier to escape from the first interfacial layer of the liquid. The free energy barriers, $\Delta A^{\dagger}$ obtained from Eqn.\ \ref{eqn:8},  are tabulated in Table \ref{tab:des}.  At a constant $\epsilon_{LL}$ the free-energy barrier increases with the increase in $\epsilon_{LS}$, which is expected because increasing the solid-liquid interaction increases the energy barrier to desorption.  However, the nature of free-energy barrier changes with the variation of liquid-liquid interaction is not as straight-forward.   With the increase in liquid-liquid interaction, the adsorbate is pulled more strongly towards the bulk liquid, which causes a decrease in the barrier. However, with the increase in liquid-liquid interaction the desorbing liquid molecule requires extra energy to make room for itself into the liquid. The latter effect will tend to increase the free-energy barrier.  These two effects work against each other and it is difficult to guess their effect on the barrier.  This is an essential difference between desorption into a liquid and desorption into a gas or vacuum.  	This behavior is similar to that observed in the photodissociation of molecules.  The quantum yield for photodissociation in a liquid is about a thousand times smaller than in the gas.  The reason for this is the so called ``cage effect'';  after photon absorption the fragments start flying apart but they hit the surrounding liquid molecules and recombine; there is no such recombination for photodissociation in a gas or in vacuum.   
 
In Table \ref{tab:des}  we give, in the third column,  the desorption energy $\langle \Delta E_{des}^{gas}$ $\rangle$ of one atom from the surface.  The surface in the supercell has 72 possible binding sites.  The ratio 1/72 in the table indicates that in the calculations reported in the third column there is one adsorbed atom in the supercell. There is therefore no lateral interaction between adsorbates.

The quantity $\langle \Delta E_{des}^{gas}$ $\rangle$ was calculated as follows.  A Monte-Carlo simulation was performed for the whole system: the solid, the interface and the liquid.  When the system has equilibrated we removed all liquid atoms from the system except those in the immediate neighbourhood of the solid surface.  We call the remaining system a full monolayer (this is the  number of atoms on the surface when the solid is in contact with the liquid film).  We then calculated the energy of desorbing one atom from a full monolayer.  This energy differs from one Monte Carlo configuration to another and this is the reason we report its mean value.   This is the energy required to desorb a molecule without the ``cage effect'' caused by the presence of the liquid film.  The surface of the solid in the simulation has 72 sites on which the liquid atoms could adsorb.  A full monolayer consists of a smaller number of molecules because of the ``lattice mismatch'' between the surface and the adsorbed layer;  the potentials we use are such that the equilibrium distance between two liquid atoms is larger than the distance between adjacent binding sites on the solid surface.
	 
The desorption energy of one atom  into the gas-phase increases with the increased coverage. For example, the desorption energy at a coverage of 40/72 (forty adsorbed atoms per 72 lattice sites)  is 46 kJ/mol, which is  $\sim$ 20 kJ/mol higher than  the desorption energy of a single adsorbed atom.  At higher coverages for the same liquid-liquid and solid-liquid interaction, the repulsive forces will reduce the desorption barriers. In general, we find that, for a given solid-liquid interaction, the energy to desorb in the gas-phase increases with the increase in liquid-liquid interaction.

We also report average energy $\langle E^{\dagger}\rangle - \langle E_{ads}\rangle$ in Table \ref{tab:des} required for moving an atom from the adsorbed layer to the second layer in the liquid.  For a given solid-liquid interaction the desorption energy from the  first to the second ad-layer of the liquid  decreases with the increase in liquid-liquid interaction. This trend is opposite to that observed in the gas-phase. Additionally, the magnitude of desorption energies are considerably lower than that observed in the gas-phase. The liquid film helps lowers the desorption barrier. This is consistent with the short lifetime of the adsorbed atoms observed in the molecular-dynamics simulations. 
  
	We also tabulate the desorption free energy barriers $\Delta A^{\dagger}$ (defined by Eqn.\ \ref{eqn:8}) and $\Delta F^{\dagger}$ which is the barrier to desorption on the potential of mean force.   Both barriers increase with the solid-liquid interaction, which is expected, and decrease when liquid-liquid interaction decreases.  We attribute the latter to the fact that as the liquid-liquid interaction decreases it is less difficult to inject the desorbing atom into the liquid.

\subsection{The desorption rate coefficient}
 We have used the Monte Carlo procedure to equilibrate the system and recorded the number $N_{a}(0)$ of atoms adsorbed on the surface after equilibration.  We then start the molecular dynamics simulation and record the number of atoms $N(t)$ that are still on the surface at time $t$.  $N(t)$ decays in time and the curve $N(t)$ versus $t$ is rather rugged.  Averaging it over 100 simulations produces a smooth curve which is fitted well by a decaying exponential (Eqn.\ \ref{eqn:2}) with $\langle N_a(t) \rangle_{\text{MC + MD}}$ calculated by Monte Carlo. We denote $\langle k \rangle$  obtained by this procedure by $k_d$ and call it the exact desorption rate constant.  The magnitude of $k_d$  is shown in Table \ref{tab:ratepar} for several values of $\epsilon_{LL}$ and $\epsilon_{LS}$.  
	
	As seen in Table \ref{tab:ratepar}, the desorption rate constant decreases when the solid-liquid interaction $\epsilon_{LS}$ is increased, which is expected. The desorption rate constant increases when $\epsilon_{LL}$ is made smaller, which is counterintuitive.  We attribute this to the increase of the energy needed to make room for the desorbing atom in the liquid.  
	
	We have  calculated the desorption rate constant for three temperatures and fitted (see Fig.\ S3 ) the result to $k­_d = \mathcal{A} \exp\left[-E_{\text{act}}/k­_BT\right]$. The pre-exponential $\mathcal{A}$ and the activation energy $E_{\text{act}}$ are given in Table \ref{tab:ratepar}.  Based on transition state theory one expects a pre-exponential of the-order-of $10^{13}$. This is not what we obtain and one speaks of anomalous pre-exponentials.   The activation energy increases when $\epsilon_{LS}$ increases, which is expected. The dependence of $E_{\text{act}}$ on $\epsilon_{LL}$ is the same as that of $k_{d}$, for the same reasons.

	We have implemented in two ways the transition state theory for the desorption rate constant.  In one we use the barrier $\Delta A^{\dagger}$ calculated with Eqn.\ \ref{eqn:8}; in the other we use the barrier energy $\Delta F^{\dagger}$ on the potential of mean force.  In the columns 7 and 9 we give the ratio of the exact rate constant $k_d$ and the rate constants calculated with transition state theory.  The rate constant obtained by using $\Delta A^{\dagger}$ for the activation energy is very different from $k_d$;  the one using $\Delta F^{\dagger}$ is much better but it still can be off by almost an order of magnitude. It is also curious that $\Delta F^{\dagger}$ is very different from the activation energy of the exact rate constant.  
	
\begin{figure}[!h] 
\includegraphics[scale=1]{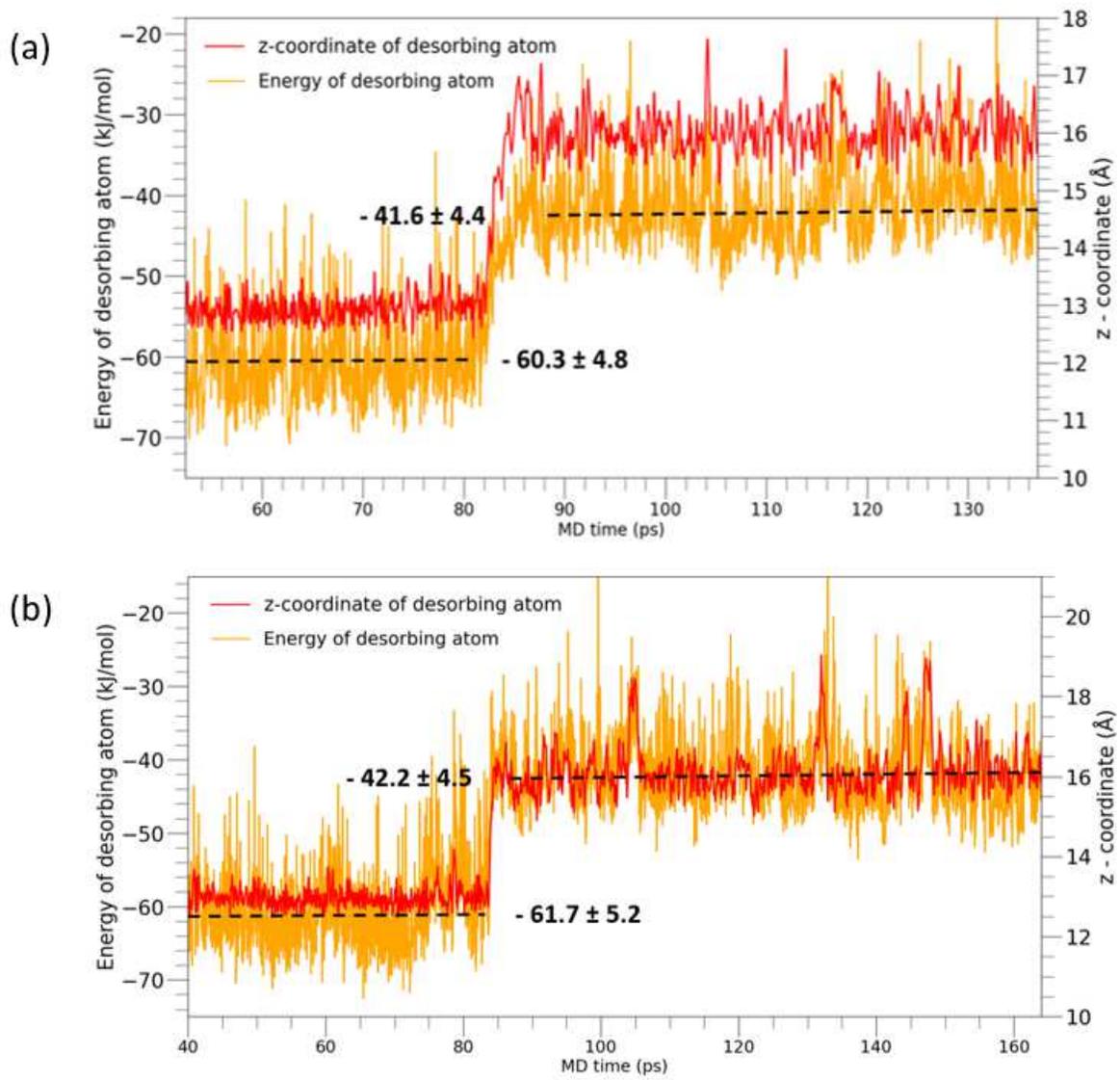}
\caption{Two sample trajectories showing energy and z-coordinate of desorbing atom. The Lennard-Jones parameters used for these plots are $\epsilon_{\text{LL}} = 3.3$ kJ/mol and $\epsilon_{\text{LS}} = 4$ kJ/mol .}
\label{fig:adsdesorp}
\end{figure}

	One source of errors in transition state theory is the assumption that once past of the dividing surface the atom does not return to the surface to readsorb. This is correct for desorption into vacuum but not for desorption into a liquid.  The second one is the neglect of fluctuations. Each atom desorbs according to its own instantaneous situation: its velocity and the positions of the atoms surrounding it. In particular, an atom that happens to have enough kinetic energy in the $z$ direction and happen to have room to go in the liquid has a very different chance to desorb then one that is caged by the liquid and is moving in the wrong direction.   In Fig.\ \ref{fig:adsdesorp} we show two trajectories of a desorbing atom.  The yellow line is the energy of the desorbing atom (this is the total interaction energy of the atom being watched with all other atoms) and the red one is the position of the same atom.  The atom starts on the surface and for a while it is trapped there:  its $z$-coordinate undergoes irregular back and forth oscillations. After about 90 picosecond the atom  leaves the surface and undergoes irregular motion inside the liquid.   These two trajectories do not return to the solid surface, but other trajectories do.  The yellow curves show the total energy of the atom being watched (this is the interaction energy of the watched atom with all the other atoms in the system).  The mean energy (averaged over 100 such trajectories) is shown by the black dashed line. The difference between the mean energies are 18.7 kJ and 19.5 kJ.  
	
	The point is that each atom desorbs according to its peculiar environment.  Replacing the real trajectories with one mean trajectory will cause errors.  

\subsection{The compensation effect}
 The compensation effect is assumed to apply to a set of similar reactions.  For example, one might study the same reaction in several solvents.  Or one might study the same oxidation reaction performed on different oxide catalysts.  Let's assume that we have studied such a set of similar reactions, determined the rate constant $k_i$ of each one of them and found that they all satisfy an Arrhenius relation $k_i = \mathcal{A}_i exp\left[ - E_i/RT\right]$.  The compensation effect states that the logarithm of the pre-exponential depends linearly on the activation energy: 	$\ln\left(\mathcal{A}_i\right) = a + b E_i$.
The proposal is controversial as can be seen from some of the review article discussing it.\cite{BOND2000}  In our system the desorption rate, the pre-exponential and the activation energies are function of $\epsilon_{LL}$ and $\epsilon_{LS}$.  If we fix the value of $\epsilon_{LS}$ and we denote them $k_d\left(\epsilon_{LS}, \epsilon_{LL}\right)$, $\mathcal{A}\left(\epsilon_{LS},\epsilon_{LL}\right)$ and $E_{\text{act}}\left(\epsilon_{LS}, \epsilon_{LL}\right)$. If the compensation effect takes place in our system then for every fixed value of $\epsilon_{LS}$ the logarithm of $\mathcal{A}$ must be a linear function of $E_{\text{act}}$.  We find that 
\begin{eqnarray}
\ln[A(\epsilon_{LL})] & = & 22.86+0.382E_{\text{act}}(\epsilon_{LL}) \hspace{10mm} \text{for} \hspace{10mm} \epsilon_{LS} = 3.0~ \text{kJ}, \\
\ln[A(\epsilon_{LL})] & = & 23.10+0.343E_{\text{act}}(\epsilon_{LL}) \hspace{10mm} \text{for} \hspace{10mm} \epsilon_{LS} = 3.5~ \text{kJ}, \\
\ln[A(\epsilon_{LL})] & = & 24.08+0.308E_{\text{act}}(\epsilon_{LL}) \hspace{10mm} \text{for} \hspace{10mm} \epsilon_{LS} = 4.0~ \text{kJ}. 
	\end{eqnarray}
While we do not have data for a large number of values of $\epsilon_{LL}$ it appears that compensation is present in our system.  
 
\section{Appendices}
\begin{appendices}
\renewcommand{\theequation}{A.\arabic{equation}}
\setcounter{equation}{0}  
\renewcommand{\thesection}{Appendix \Alph{section}}
\section{The Computation of Probability Distributions}
We compute $P(\xi)$ by dividing the liquid into bins (of width $\Delta_{b}$) defined by planes parallel to the solid surface.  We take $\Delta_{b}$ to be much smaller than $z$ so that the free energy is constant within each bin. We then create a histrogram by performing Monte Carlo simulations. Let $n_i$ be the number of particles whose center of mass is in bin $i$, then $P(\xi)$ can be obtained by performing a very large number of Monte Carlo moves ($N_{MC}$) and is given by  
\begin{equation}
P(\xi) \Delta_{b} = n_{i}/ N_{MC}.
\end{equation}
A Metropolis implementation of Monte Carlo method is usually not sufficient to sample high energy configurations. This is a rare-event problem which can be solved by using a multiple-window umbrella sampling method.\cite{Torrie1974, Kumar1992} Umbrella sampling uses a bias potential, which discourages the system to wander very far away from the minima of the bias potential. In this work, we use the bias potential of the following form
 \begin{equation}
 U_{j}(\xi)=\frac{1}{2}k_{j}(\xi-\xi_{j})^2 \hspace{10mm} j = 1, 2, ...,N_{w},
 \end{equation}
where $N_w$ is the number of umbrella windows and $\xi_{­j} = \xi_1 + j\Delta W$. 
We obtain the unbiased distributions by combining the data from each  ``umbrella window'' using the weighted histogram analysis method (WHAM) \cite{Kumar1992, Gallicchio2005}  in which the unbiased distribution $P^i$ is obtained from 
\begin{equation}
P^i = \frac{\Sigma_{j=1}^{N_W} {n_{ij}}}{\Sigma_{j=1}^{N_W} {N_{j} \exp[-U_j(\xi_j)/k_{B}T]/\mu_j}}.
\end{equation}	  				
Here $n_{ij}$ is the number of visits in the $i$-th bin and $j$-th window, $N_w$ is the number of windows,  $N_b$ is the number of bins, $N_{j}=\Sigma_{i=1}^{N_B} n_{ij}$ and $\mu_{j}=\Sigma_{i=1}^{N_B} \prod_{i} \exp[-U_j(\xi_j)/k_{B}T]$. The equations are solved self-consistently by providing an initial guess of $\mu_{j}$.

For each umbrella window the system was equilibrated for $10^6$ sweeps.  Each sweep consists off as many Monte Carlo moves as particle in the system. Data was collected during each sweep and a total of $2.0\times10^6$ sweeps were performed in each umbrella window for getting the potential of mean force. No appreciable change in the potential of mean force was found when performing an additional $10^6$ sweeps. The width of each umbrella window was 0.1 \AA\ and the constant in the bias potential was $k_j = 800 $ kJ/mol \AA$^2$.  These values were decided by performing several benchmarking runs. The probability distribution was calculated with a bin width $\Delta_b$ = 0.01 \AA.

\end{appendices}

\begin{acknowledgments}
Krishna Jaiswal acknowledges support from MHRD, Govt.\ of India for PhD scholarship. Vishal Agarwal acknowledges financial support from Science and Engineering Board (SERB), Department of Science and Technology, Govt.\ of India (Ramanujan Fellowship).
The authors acknowledge the support of DST, Govt.\ of India, for the High-Performance Computing (HPC) facility  at IIT Kanpur.
The authors also acknowledge the NSM (National Supercomputing Mission) facility, Govt.\ of India, for the Param Sanganak facility at IIT Kanpur.
\end{acknowledgments}
\clearpage

\bibliography{new_ref}

\clearpage

\clearpage

\clearpage


\clearpage

\clearpage

\begin{table} 
\caption{\textbf{Model parameters for pairwise interaction.}}
\label{tab:parm}
\centering
\begin{tabular}{p{1.2cm} p{1.2cm} p{1.8cm} p{2cm}} 
\hline\hline 
\multicolumn{4}{l}{\textbf{The parameters in the Morse Potential.}} \rule{0pt}{3ex} \\
\cline{2-4}
& $r_e$ (\AA) & $\alpha$ (\AA$^{-1}$) &  $D_e$ (kJ/mol)\\
\cline{2-4}
S-S & 2.838  &1.4 & 33 \\    
                \\
 \multicolumn{4}{l}{\textbf{The parameters in the Lennard-Jones Potential. }}            \\
 \cline{2-3}
& $\sigma$ (\AA) & $\epsilon$ (kJ/mol) &  \\
\cline{2-3}
   S-L   & 3.3   &  3, 3.5, 4      &  \\  
   L-L   &  3.2 &   3, 3.3, 3.6 &    \\
\hline\hline
\end{tabular}
\end{table}

\clearpage
\begin{table} 
\caption{\textbf{Desorption energy ($\Delta E^{\text{gas}}_{\text{des}}$) of an atom from the surface to gas-phase as a function of Lennard-Jones parameters. The number in parenthesis represents the initial coverage on the surface. The monolayer coverage was computed from the average number of atoms when a liquid layer was present. $\langle E^{\dagger} \rangle$  - $\langle E_{\text{ads}} \rangle$ is the difference in energy of an desorbing atom from the first ad-layer to the second ad-layer. $\Delta F^\dagger$ is the restricted free energy barrier computed from the potential of mean force. $\Delta A^\dagger$ is the Helmoltz free energy barrier computed using Eqn.\ \ref{eqn:8}. }}
\label{tab:des}
\centering
\begin{tabular}{>{\centering\arraybackslash}p{1cm} >{\centering\arraybackslash}p{1cm} >{\centering\arraybackslash}p{2.1cm} >{\centering\arraybackslash}p{2.2cm}  
>{\centering\arraybackslash}p{2.8cm} >{\centering\arraybackslash}p{2cm} 
>{\centering\arraybackslash}p{2cm}}
\hline\hline 
\\[-0.8em]
\multicolumn{2}{l}{$\epsilon$ (kJ/mol)} & $\Delta E^{\text{gas}}_{\text{des}}$ &   $\langle\Delta E^{\text{gas}}_{\text{des}}\rangle$ (monolayer) & $\langle E^{\dagger} \rangle$  - $\langle E_{\text{ads}} \rangle$ & $\Delta F^\dagger$  &  $\Delta A^\dagger$ \\
$\epsilon_{\text{LS}}$ & $\epsilon_{\text{LL}}$ & (kJ/mol) & (kJ/mol) & (kJ/mol) & (kJ/mol) & (kJ/mol) \\
\\[-0.8em]
\hline
\\[-0.8em]
\multirow{
3}{*}{3} & 3.6 & 28 (1/72) & 46 (40/72)  & 10 & 20 & 29\\
\\[-0.8em]
& 3.3 &  28 (1/72) & 43 (39/72) & 12 & 15 & 25\\
\\[-0.8em]
& 3 & 28 (1/72) & 38 (39/72) & 13 & 13 & 23\\
\\[-0.8em]
\hline
\\[-0.8em]
\multirow{
3}{*}{3.5} & 3.6 &  33 (1/72) & 52 (42/72) & 12 & 23 & 33\\
\\[-0.8em]
& 3.3 &  33 (1/72) & 50 (42/72) & 16 & 19 & 29 \\
\\[-0.8em]
& 3 &  33 (1/72) & 49 (41/72)& 17 & 16 & 26 \\
\\[-0.8em]
\hline
\\[-0.8em]
\multirow{
2}{*}{4} & 3.6 &  38 (1/72) & 58 (42/72) & 18 & 29 & 38\\
\\[-0.8em]
& 3.3  & 38 (1/72) & 56 (42/72) & 19 & 24 & 33\\
\\[-0.8em]
& 3 & 38 (1/72) & 52 (41/72) & 20 & 20 & 29\\
\\[-0.8em]
\hline\hline
\end{tabular}
\end{table}

\begin{table} 
\caption{\textbf{The exact rate of desorption $k_d$. $\mathcal{A}$ and $E_{\text{act}}$ are the pre-exponential and the activation energy obtained by fitting the temperature dependence of $k_d$ with the Arrhenius formula. $\Delta F^\dagger$ is the restricted free energy barrier computed from the potential of mean force. $\Delta A^\dagger$ is the barrier computed with Eqn.\ \ref{eqn:8}. The values in the parenthesis represents the power of 10. The columns 7 and 9 give the ratio of the exact desorption rate constant to the value calculated with transition state theory by using the mean activation free energies $\Delta A^\dagger$ and $\Delta F^\dagger$, respectively.}}
\label{tab:ratepar}
\centering
\small
\begin{tabular}{>{\centering\arraybackslash}p{1cm} >{\centering\arraybackslash}p{1cm} >{\centering\arraybackslash}p{1.7cm} >{\centering\arraybackslash}p{1.7cm} >{\centering\arraybackslash}p{1.7cm} >{\centering\arraybackslash}p{1.7cm}
 >{\centering\arraybackslash}
p{2.4cm} >{\centering\arraybackslash}p{1.7cm}
>{\centering\arraybackslash}p{2.4cm}
} 
\hline\hline 
\\[-0.8em]
\multicolumn{2}{l}{$\epsilon$ (kJ/mol)} &  $k_d$ & $\mathcal{A}$ &  $E_{\text{act}}$   &  $\Delta A^\dagger$ &  \multirow{
2}{*}{$\frac{k_d}{\frac{k_BT}{h}\exp\left(-\frac{\Delta A^\dagger}{RT}\right)}$} & $\Delta F^\dagger$   & \multirow{
2}{*}{$\frac{k_d}{\frac{k_BT}{h}\exp\left(-\frac{\Delta F^\dagger}{RT}\right)}$}  \\
$\epsilon_{\text{LS}}$ & $\epsilon_{\text{LL}}$ & (s$^{-1}$) & (s$^{-1}$)  & (kJ/mol) & (kJ/mol) &  & (kJ/mol) & \\
\\[-0.8em]
\hline
\\[-0.8em]
\multirow{
3}{*}{3} & 3.6 & 4.28 (9) & 7.33 (14) & 31 & 29 & 83.2 & 20 & 1.80 \\
\\[-0.8em]
& 3.3 & 4.98 (9) & 1.33 (14) & 26 & 25  &  17.0 & 15 & 0.34 \\
\\[-0.8em]
& 3 & 8.31 (9) & 2.20 (13) & 20 & 23  & 12.0 & 13 & 0.23 \\
\\[-0.8em]
\hline
\\[-0.8em]
\multirow{
3}{*}{3.5} & 3.6 & 6.13 (8) & 8.31 (17) & 52 & 33 & 55.7 & 23 & 1.14 \\
\\[-0.8em]
& 3.3 & 1.34 (9) & 5.31 (15) & 40 & 29  & 22.3 & 19 &  0.48 \\
\\[-0.8em]
& 3 & 2.57 (9) & 2.20 (14) & 28 & 26  & 14.2 & 16 &  0.29\\
\\[-0.8em]
\hline
\\[-0.8em]
\multirow{
2}{*}{4} & 3.6 & 8.53 (7) & 1.14 (19) & 64 & 38  &  56.8  & 29 & 1.56\\
\\[-0.8em]
& 3.3  & 1.34 (8) & 1.58 (17) & 51 & 33  & 13.4  & 24 & 0.28\\
\\[-0.8em]
& 3 & 1.85 (8) & 1.85 (16) & 43 & 29  & 12.6 & 20 &  0.27\\
\\[-0.8em]
\hline\hline
\end{tabular}
\end{table}

\end{document}